\documentstyle[epsfig]{aipproc}

\newcommand{\gap}{\mathrel{ \rlap{\raise.5ex\hbox{$>$}}
                      {\lower.5ex\hbox{$\sim$}}  } }
\newcommand{\lap}{\mathrel{ \rlap{\raise.5ex\hbox{$<$}}
                      {\lower.5ex\hbox{$\sim$}}  } }  

\begin{document}
\title{Thermonuclear Runaways on Accreting White Dwarfs: Models of Classical 
Novae Explosions}
\author{Margarita Hernanz and Jordi Jos\'e}
\address{Institute for Spatial Studies of Catalonia (IEEC/CSIC/UPC),
Edifici Nexus-201, \\
C/Gran Capit\`a 2-4, 08034 Barcelona (SPAIN).}
\maketitle

\begin{abstract}  
The mechanism of classical novae explosions is explained, together with some
of their observational properties. The scarce but not null impact of novae
in the chemical evolution of the Milky Way is analyzed, as well as their
relevance for the radioactivity in the Galaxy. A special emphasis is given to
the predicted gamma-ray emission from novae and its relationship with the
thermonuclear model itself and its related nucleosynthesis.
\end{abstract}

\section*{Introduction}
Classical novae are a very common type of cosmic explosion. They occur very
often in the Galaxy ($\sim$ 35 yr$^{-1}$), although only $\sim$3-5 are
discovered by amateur astronomers in the Galaxy every year.
The explosion occurs on the top of white dwarfs accreting
mass in a cataclysmic binary system and it is related to degenerate hydrogen
ignition, leading to a thermonuclear runaway. Hydrogen burning occurs mainly
through the CNO cycle, which operates out of equilibrium, because
radioactive nuclei of lifetimes longer than the evolutionary timescale are
synthesized. Convection also plays a crucial role, since it transports some
of the $\beta^+$-unstable nuclei synthesized to the outer envelope, where their
subsequent decay powers the expansion of the envelope and the increase of
luminosity.

Although classical novae release large amounts of energy ($\sim 10^{45}$ erg),
they don't have an impact in the interstellar medium dynamics, like supernovae.
Concerning the chemical evolution of the Galaxy, they account only for
$\sim 1/3000$ of the Galactic disk's gas and dust content; therefore, novae
scarcely contribute to Galactic abundances, except for some particular elements,
like $^{13}$C and $^{17}$O. Also, novae contribute to the radioactivity of
the Galaxy, through the emission of $\gamma$-rays related to the decay of
some medium and long-lived nuclei (i.e., $^{7}$Be, $^{22}$Na and $^{26}$Al).

Novae have been observed in all energy ranges, except in the $\gamma$-ray domain,
where current instruments are not sensitive enough. From the observations in
the optical, ultraviolet and infrared, information about the physical conditions
of the ejecta and its composition has been deduced. On the other hand, soft X-ray
observations, when available, have provided some information about the turn-off
of novae. A good example of a nova observed at all wavelengths is Nova Cyg 1992.

This paper has been organized as follows: first of all, some observational
properties of classical novae are presented, mainly the light curves. Then, the
thermonuclear runaway model is explained and illustrated with some numerical
results. Finally, nucleosynthesis in classical novae and its relevance for the
chemical evolution and the radioactivity of the Galaxy, with a spetial mention
to theoretical models of $\gamma$-ray emission from novae, is presented. 

\section*{Some observational properties of novae}
Visual light curves of classical novae have some general trends that
distinguish them from other variable stars, such as supernovae or dwarf novae,
to cite two extreme examples. First of all, there is an increase in luminosity
which corresponds to a decrease of m$_{\rm V}$ (apparent visual magnitude) 
of more than 9 magnitudes occuring in a few days. In some cases, a pre-maximum
halt, 2 magnitudes before maximum, has been observed (see \cite{Wa95} and
references therein). 

In order to characterize the nova light curves, their speed class is defined
from either t$_2$ (or t$_3$), which is the time needed to decay in 2 (or 3)
visual magnitudes after maximum. Novae speed classes range from very fast
(t$_2 < 10 $ days) and fast (t$_2 \sim 11-25$ days) to very slow 
(t$_2 \sim 151-250$ days) \cite{Pa57}.
Nova Cyg 1992, for instance, had t$_2 \sim 12$ days, which is quite fast,
and Nova Her 1991 was even faster (t$_2 \sim 2$ days). An example of a slow
nova is Nova Cas 1993, which had t$_2 \sim 100$ days. 

There is a relationship between the absolute magnitude at maximum M$_{\rm V}$ 
and the speed class of novae, in the sense that brighter novae have shorter
decay times (t$_2$ or t$_3$). The theoretical
explanation of this relationship \cite{Li92} is based on the widely accepted 
model of nova explosions, which shows that novae reach a luminosity at maximum
which is close to the Eddington luminosity, and also that novae should eject
roughly all their envelope in a time similar to t$_3$. Thus, one can establish
quantitatively that L$_{\rm max}$ is an increasing function of M$_{\rm wd}$ 
and that t$_3$ is a decreasing function of M$_{\rm wd}$. From these two 
relationships a new one can be obtained which relates  M$_{\rm V}$ at maximum 
with t$_3$. This empirical relation (which is valid both in the V and B 
photometric bands) is very often used to determine distances to novae, once
visual extinction is known. Different calibrations of the
maximum magnitude-rate of decline relationship (MMRD) exist, with that
from \cite{DL95} being the most usual one (see also \cite{Sh97}). 

It is important to mention that it is in general assumed that two different
kinds of nova populations exist \cite{Du90}: the disk population (with
scaleheight z$\lap$ 100 pc) made of bright and fast novae, and the bulge
population (with scaleheight up to z$\gap$ 1 kpc), made of dimmer and slower
novae \cite{DV92}. Also, Della Valle \& Livio \cite{DL98} have
established a link between these disk and thick-disk/bulge novae and the
spectroscopic classification of Williams \cite{Wi92}.

Since the launch of astronomical satellites and, specially, of the IUE
(International Ultraviolet Explorer), light curves have been extended to other
energetic domains away from the optical one. It was discovered from IUE
observations of novae that the luminosity in the ultraviolet band increases 
when the
optical one starts to decline: the reason is that there is a shift of the
energy distribution to higher energies, because deeper and hotter regions of
the expanding envelope are seen (the photosphere recedes because 
opacity decreases when temperature falls below 10$^4$K, the recombination 
temperature of hydrogen). Also, infrared observations when available (i.e.,
for novae in which dust forms) indicate 
an increase once the ultraviolet luminosity starts to decline, which is 
interpreted as
the resulting reradiation by dust grains in the infrared of the ultraviolet 
energy they have absorbed. In summary, the
bolometric luminosity of classical novae is constant for a quite long period
of time, being the duration of this constant L$_{\rm bol}$ phase dependent on
the remaining envelope mass of the nova.

The constancy of L$_{\rm bol}$ has been interpreted
and obtained theoretically, although the concomitant mass-loss has not been 
well understood and modeled. The phase of constant L$_{\rm bol}$ corresponds
to hydrostatic hydrogen burning in the remaining envelope of the nova,
accompanied by a continuous mass-loss probably by an optically thick wind.
Since the bolometric luminosity deduced from observations is close to or even 
larger than the Eddington luminosity, radiation pressure is probably the main 
force causing ejection of nova envelopes. An additional observational proof 
of this phase has
come from the observations in the soft X-ray range. The EXOSAT satellite
detected the nova GQ Mus (Nova Mus 1983) as a soft X-ray emitter (in the
interval 0.04-2 keV), 460 days after optical maximum \cite{Og84}. ROSAT
detected again that source (0.1-2.4 keV), even 9 years after the explosion
\cite{Og93}. Nova Cyg 1992 was detected by ROSAT too as a powerful soft
X-ray source, but the emission lasted in that case only for one year and a
half \cite{Kr96}. The interpretation of the
soft X-ray emission is that it is related to blackbody emission of the
remaining hydrogen-burning shell, which becomes visible when the expanding
envelope is transparent to it. The luminosity deduced is close to 
L$_{\rm Edd}$, thus indicating again the constancy of L$_{\rm bol}$ and the
hardening of the spectra. It is worth mentioning that in some novae
(Nova Her 1991, Nova Pup 1991, Nova Vel 1999) hard X-ray
emission has also been detected, with much smaller luminosities \cite{Ll92}.
In these cases, the interpretation is different, since the emission mechanism
is probably related to bremsstrahlung in some shocked region around the nova. 

The observed turn-off times of novae, deduced from soft X-ray and ultraviolet 
observations, are between 1 and 5 yr \cite{GR98}, 
except for Nova Mus 1983 (9-10 yr). This is much shorter than expected from
the nuclear burning timescale of the remaining envelope, thus telling that some
extra mechanism besides of nuclear reactions leads to the extinction of the 
shell.  

Spectra of novae are quite complicated and show different features related to
some particular phases: four succesive systems of absorption lines, and
five overlapping systems of emission lines are present in almost all novae.
From the emission-line spectra in the nebular phase, both in the optical and
the ultraviolet, and also from infrared spectra in some cases, detailed
abundance determinations of nova ejecta are available (see \cite{Ge98}
for a recent review). A general trend is observed: in many novae there is an
enhancement of metallicities above solar and, in particular, enhancements of
carbon-nitrogen-oxygen (CNO) elements and/or neon. It is known since long
ago (see for instance \cite{St78a} and \cite{Pr78}) that
some enrichment of the accreted matter (which is in principle assumed to be of
solar composition) with the underlying white dwarf core (of the CO or ONe type)
is necessary, both to power the nova explosion and to explain the observed
enhancements (see the following section for details).

\section*{Thermonuclear runaway model of nova explosions}
The accepted scenario of classical novae explosions is the thermonuclear
runaway model, in which a cold white dwarf in a cataclysmic variable accretes
hydrogen-rich matter, as a result of Roche lobe overflow of the main sequence
companion. If the accretion rate is low enough (e.g., $\dot{M} \sim
10^{-9}-10^{-10}$ M$_\odot$ yr$^{-1}$), 
the accreted hydrogen is compressed up to degenerate conditions,
thus leading to thermonuclear burning without control (thermonuclear runaway).
The explosive burning of hydrogen produces some $\beta^+$-unstable nuclei of
relatively short timescales (i.e., $^{13}$N, $^{14}$O, $^{15}$O, $^{17}$F),
which are transported by convection to the outer envelope, where they are
preserved from destruction until they decay. Their subsequent decay implies a
huge liberation of energy in the outer shells, which originates envelope
expansion, increase in luminosity and mass ejection, as required in a classical
nova explosion. 

In order to understand the thermonuclear runaway (TNR) mechanism for nova
explosions, it is important to evaluate some relevant timescales (see
\cite{St89} for a review). The accretion timescale, defined as
$\tau_{\rm acc} \sim {\rm M_{acc} / \dot{M}}$ (which is of the order of
$10^4-10^5$ yr, depending on the accretion rate $\dot{M}$), the nuclear
timescale $\tau_{\rm nuc} \sim {\rm C_p T / \epsilon_{nuc}}$ (which is as
small as some seconds at peak burning), and the dynamical timescale
($\tau_{\rm dyn} \sim {\rm H_p / c_{s} \sim (1/g) \sqrt{P/\rho}}$). During the
accretion phase, $\tau_{\rm acc} \le \tau_{\rm nuc}$, accretion can
proceed and increase the envelope mass. When degenerate ignition conditions
are reached, degeneracy prevents envelope expansion and the TNR occurs. As
temperature increases, degeneracy would be lifted (since T would become larger
than T$_{\rm Fermi}$) and expansion would turn-off
the explosion, but this is not so because $\tau_{\rm nuc} \ll \tau_{\rm dyn}$
(specially if the envelope is enriched in CNO elements above solar values,
thus enhancing the contribution of the CNO cycle to hydrogen burning).
Therefore, since the envelope can not respond by expanding, temperature
and nuclear energy generation rate, $\epsilon_{\rm nuc}$, continue to increase 
without control. The value of the nuclear timescale is crucial for the 
development of the TNR and its final fate. In fact there are mainly two types
of nuclear timescales: those related to $\beta$-decays, $\tau_{\beta^+}$, and
those related to proton capture reactions, $\tau_{(\rm p,\gamma)}$. In the 
early evolution towards the TNR, $\tau_{\beta^+} < \tau_{(\rm p,\gamma)}$ and 
the CNO cycle operates in equilibrium. But as temperature increases up to
$\sim 10^8$ K, the reverse situation is true
($\tau_{\beta^+} \gap \tau_{(\rm p,\gamma)}$), and thus the CNO cycle is
$\beta$-limited. In addition, since the large energetic output produced by
nuclear reactions can't be evacuated only by radiation, convection sets in and
transports the $\beta^+$-unstable nuclei to the outer cooler regions where they
are preserved from destruction and where they 
will decay later on ($\tau_{\rm conv} \lap \tau_{\beta^+}$), leading to
envelope expansion, increase in luminosity and final mass ejection if the
attained velocities are large enough. Another important effect of convection is
that it transports fresh unburned material to the burning shell. In summary,
non-equilibrium burning occurs and the resulting nucleosynthesis will be far 
from that of hydrostatic hydrogen burning. 

\begin{table}[t] 
\caption{Some properties of the explosions of CO and ONe novae}
\begin{tabular}{ccccc}
Nova type & M$_{\rm wd}$ (M$_\odot$) & Peak temperature (K)   
          & Kinetic energy of the ejecta 
          &M$_{\rm ej}$ (M$_\odot$) \\
\hline
CO        & 1.15               & $2.05 \times 10^{8}$ 
          & $1.1 \times 10^{45}$
          & $1.3 \times 10^{-5}$ \\
ONe       & 1.15               & $2.31 \times 10^{8}$ 
          & $1.5 \times 10^{45}$
          & $2.6 \times 10^{-5}$  \\
ONe       & 1.25               & $2.51 \times 10^{8}$ 
          & $1.5 \times 10^{45}$
          & $1.8 \times 10^{-5}$   \\
\end{tabular}
\end{table}

In table 1 we show some general properties of computed models, corresponding
two carbon-oxygen (CO) and oxygen-neon (ONe) novae, with accretion rate $2
\times 10^{-10}$ M$_\odot$ yr$^{-1}$. More details about these models are
given in \cite{JH98}. Other recent detailed theoretical models 
of CO and ONe nova explosions are shown in \cite{PK95} and \cite{St98}, 
respectively. 

\begin{table}[t] 
\caption{Comparison between some models of CO and ONe novae and some
observations}
\begin{tabular}{ccccccccc}
                         &    &    &    & Element&    &    &       &     \\
\cline{2-9}
Model                    & H  & He & C    & N   & O   & Ne & Na-Fe & Z   \\
\cline{1-9}
V693 CrA 1981            &    &    &      &     &     &    &       &     \\
\cline{1-9}
Vanlandingham et al. 1997&0.25&0.43&0.025 &0.055&0.068&0.17& 0.058 & 0.32\\
ONe, 1.15M$_\odot$, mixing 50\%
                         &0.30&0.20&0.051 &0.045&0.15 &0.18& 0.065 & 0.50\\
\cline{1-1}
Andre\"a et al. 1994     &0.16&0.18&0.0078&0.14 &0.21 &0.26& 0.030 & 0.66\\
ONe, 1.15M$_\odot$, mixing 75\%
                         &0.12&0.13&0.049 &0.051&0.28 &0.26& 0.10  & 0.75\\
\cline{1-1}
Williams et al. 1985     &0.29&0.32&0.0046&0.080&0.12 &0.17& 0.016 & 0.39\\
ONe, 1.25M$_\odot$, mixing 50\%
                         &0.28&0.22&0.060 &0.074&0.11 &0.18& 0.071 & 0.50\\
\cline{1-9}
V1370 Aql 1982           &    &    &      &     &     &    &       &     \\
\cline{1-9}
Andre\"a et al. 1994    &0.044&0.10&0.050 &0.19 &0.037&0.56& 0.017 & 0.86\\
ONe, 1.35M$_\odot$, mixing 75\%
                         &0.073&0.17&0.051 &0.18 & 0.14&0.24& 0.14  & 0.76\\
\cline{1-1}
Snijders et al. 1987    &0.053&0.088&0.035&0.14 &0.051&0.52& 0.11  & 0.86\\
ONe, 1.35M$_\odot$, mixing 75\%
                        &0.073&0.17&0.051 &0.18 & 0.14&0.24& 0.14  & 0.76\\
\cline{1-9}
QU Vul 1984              &    &    &      &     &     &    &       &     \\
\cline{1-9}
Austin et al. 1996       &0.36&0.19&      &0.071& 0.19&0.18& 0.0014& 0.44\\
ONe, 1.0M$_\odot$, mixing 50\%
                         &0.32&0.18& 0.030&0.034& 0.20&0.18& 0.062 & 0.50\\
\cline{1-1}
Saizar et al. 1992       &0.30&0.60&0.0013&0.018&0.039&0.040&0.0049& 0.10\\
ONe, 1.15M$_\odot$, mixing 25\%
                         &0.47&0.28& 0.041&0.047&0.037&0.090&0.0035& 0.25\\
\cline{1-9}
PW Vul 1984              &    &    &      &     &     &    &       &     \\
\cline{1-9}
Andre\"a et al. 1994     &0.47&0.23&0.073 & 0.14&0.083&0.0040&0.0048&0.30\\
CO, 1.15M$_\odot$, mixing 25\%
                         &0.47&0.25&0.073 &0.094& 0.10&0.0036&0.0017&0.28\\
\cline{1-9}
V1688 Cyg 1978           &    &    &      &     &     &    &       &     \\
\cline{1-9}
Andre\"a et al. 1994     &0.45&0.22&0.070 &0.14 & 0.12&      &      &0.33\\
CO, 1.15M$_\odot$, mixing 25\%
                         &0.47&0.25&0.073 &0.094& 0.10&0.0036&0.0017&0.28\\
\cline{1-1}
Stickland et al. 1981   &0.45&0.23&0.047 &0.14 & 0.13&0.0068&      &0.32\\
CO, 0.8M$_\odot$, mixing 25\%
                        &0.51&0.21&0.048 &0.096& 0.13&0.0038&0.0015&0.28\\
\end{tabular}
\end{table}

\section*{Nucleosynthesis in nova explosions}
The main goal of studies of nucleosynthesis in novae is, of course, to
reproduce the observed abundances in novae ejecta. Although both from the
observational and the theoretical side some uncertainties exist (different
determinations of observed abundances or uncertain initial conditions for
theoretical models), a quite good fit is obtained in many cases (see table 2,
and more details in \cite{JH98}, together with the quoted papers in the table 
for the analyses of the observations and derivation of the abundances).

\begin{figure} 
\setlength{\unitlength}{1cm}
\begin{picture}(15,10)
\put(1,0){\makebox(8,10){\epsfxsize=9cm \epsfbox{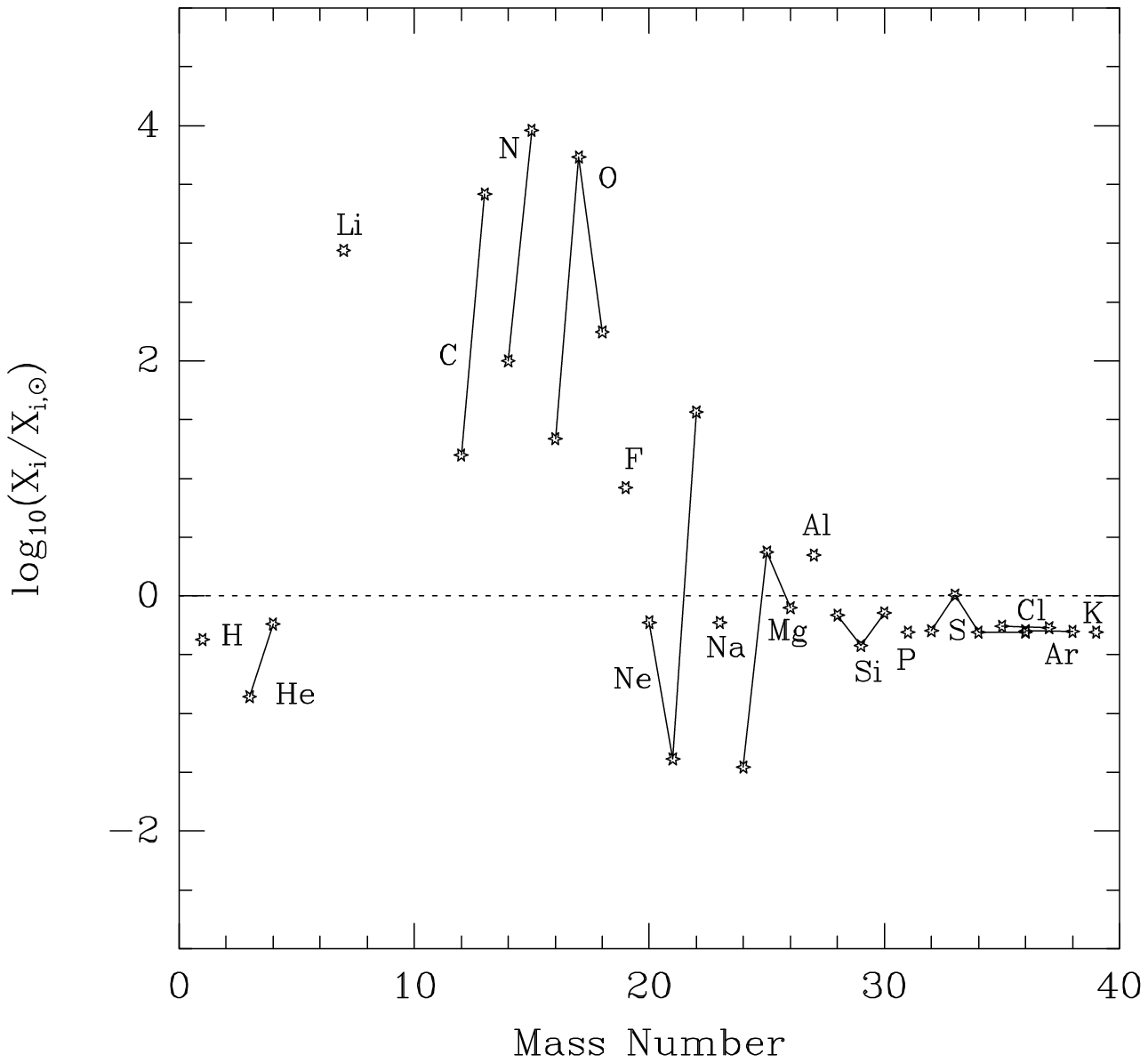}}}
\put(7.5,0){\makebox(8,10){\epsfxsize=9cm \epsfbox{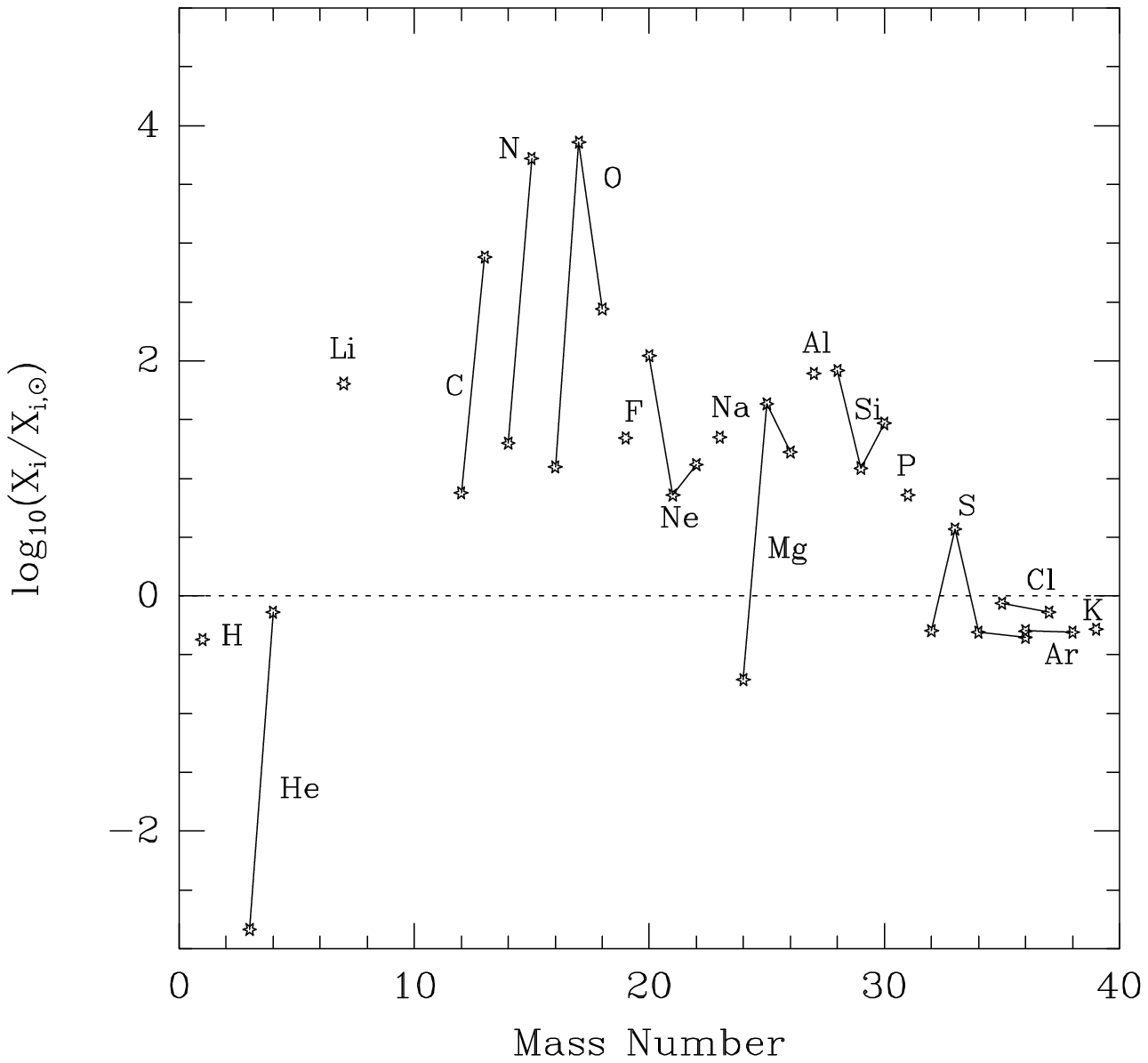}}}
\end{picture}
\vspace{-1cm}
\caption{Overproduction factors, with respect to solar abundances, obtained
for two nova models: a CO nova of 1.15 M$_\odot$ with 50\% mixing between
accreted matter and core material (left), and an ONe nova of the same mass and
degree of mixing (right).}
\end{figure}

\begin{figure} 
\setlength{\unitlength}{1cm}
\begin{picture}(15,10)
\put(1,0){\makebox(8,10){\epsfxsize=9cm \epsfbox{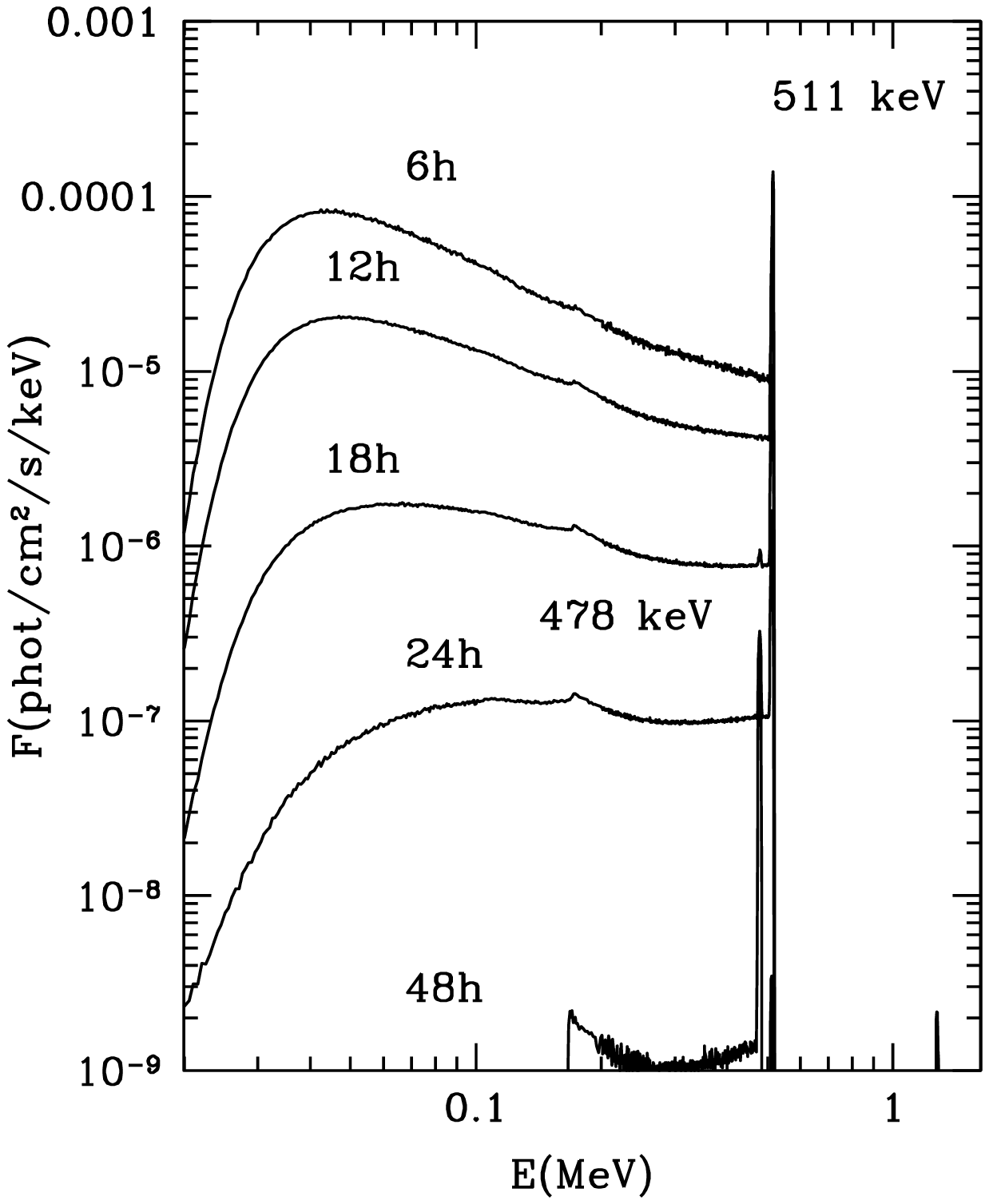}}}
\put(7.5,0){\makebox(8,10){\epsfxsize=9cm \epsfbox{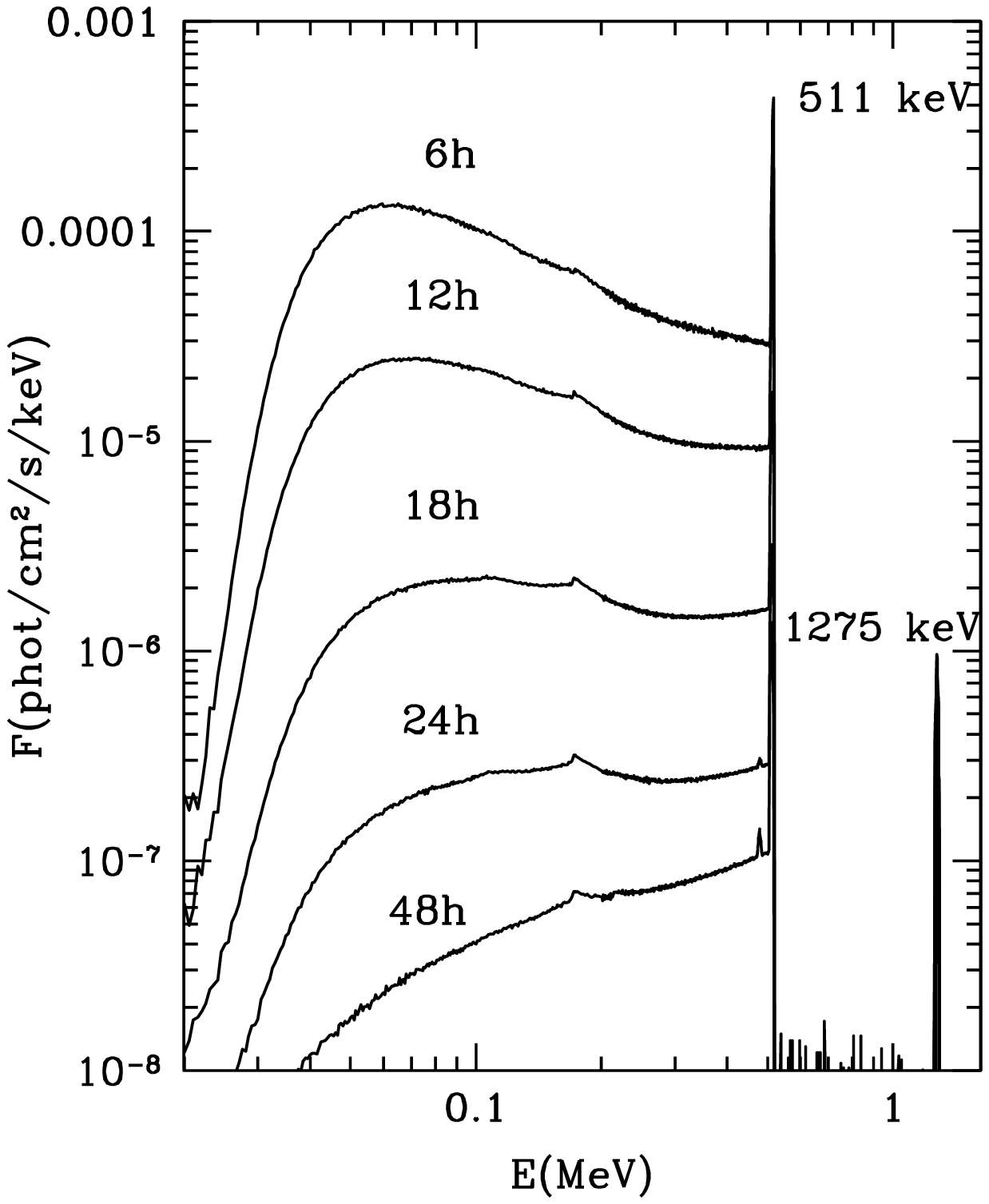}}}
\end{picture}
\vspace{-2cm}
\caption{Temporal evolution of the $\gamma$-ray emission of a CO (left) and
an ONe (right) nova, of 1.15M$_\odot$, at a distance of 1 kpc.}
\end{figure}

\subsection*{Chemical evolution of the Galaxy}
In contrast with
supernovae, novae are not important contributors to the abundances observed in
the interstellar medium, but it is also true that they can contribute to
Galactic abundances in some particular cases, when the overproduction factors
with respect to solar abundances are larger than around $10^3$ (see
\cite{JH98} and \cite{Ge98}). In figure 1 we show the
overproduction factors relative to solar abundances versus mass number for
two typical novae: a CO and an ONe ones, with mass 1.15 M$_\odot$,
50\% of mixing with core material and accretion rate
$2\times 10^{-10}$ M$_\odot$ yr$^{-1}$. In this figure some general features
are distinguishable: both in CO and in ONe novae, the largest yields correspond
to elements of the CNO group, whether in CO novae Li is also largely
overproduced. In the case of more massive ONe novae (i.e. 1.35 M$_\odot$
\cite{JH98}), intermediate-mass elements (such as Ne, Na, Mg, S, Cl) are also
overproduced.

The origin of Galactic lithium ($^7$Li) is still not completely understood.
Although it is widely accepted that there is some primordial lithium produced
during the big bang, and, of course, that spallation reactions by cosmic rays
in the interstellar medium or in flares also produce it, some extra stellar
source of $^7$Li
(without generating $^6$Li) has to be invoked. The synthesis of $^7$Li in
classical novae, by the {\it beryllium transport} mechanism \cite{Ca55},
can produce large amounts of $^7$Li, but complete hydrodynamical models are
needed in order to compute correctly the yields \cite{St78b,He96}.
$^7$Li  formation is favored in CO novae,
with respect to ONe ones (see figure 1). The reason is that CO novae
evolve faster (because
of their larger $^{12}$C content), allowing photodisintegration of $^8$B
through $^8$B($\gamma$,p)$^7$Be to prevent the destruction of the $^7$Be
synthesized during the first part of the TNR (by means of
$^3$He($\alpha,\gamma)^7$Be). Large overproduction factors with respect to
solar abundances are obtained (see figure 1), but classical novae can only
account for roughly $\sim 10 \%$ of the global Galactic $^7$Li \cite{He96}.
It is important to stress that Romano et al. \cite{Ro99} have
obtained (using our nova yields in a complete model of
chemical Galactic evolution) that the contribution from novae is
required in order to reproduce the shape of the growth of Li abundance
versus metallicity.

Concerning the nuclei of the CNO group, the main isotopes produced in novae
are $^{13}$C, $^{15}$N and $^{17}$O. The Galactic $^{17}$O is most probably
almost entirely of novae origin \cite{JH98}. Novae also contribute
significantly to the Galactic $^{13}$C and $^{15}$N, but an extra source of
$^{15}$N is required.

\begin{table}[t] 
\caption{Main radioactive isotopes ejected by novae}
\begin{tabular}{ccccc} 
Isotope   & Lifetime       &  Main disintegration process
	  & Type of $\gamma$-ray emission  & Nova type\\
\hline
$^{13}$N  & 862~s          & $\beta^+$--decay
          & 511~keV line \& continuum      & CO and ONe\\      
$^{18}$F  & 158~min        & $\beta^+$--decay
          & 511~keV line \& continuum      & CO and ONe\\
$^{7}$Be  & 77~days        & $e^-$--capture
          & 478~keV line                   & CO\\
$^{22}$Na & 3.75~years     & $\beta^+$--decay 
          & 1275~keV \& 511~keV lines      & ONe\\
$^{26}$Al & 10$^{6}$~years & $\beta^+$--decay
          & 1809~keV \& 511~keV lines      & ONe\\
\end{tabular}
\end{table}

\begin{table}[t] 
\caption{Ejected masses (in M$_\odot$) of radioactive nuclei obtained from
theoretical models of CO and ONe novae ($^{13}$N and $^{18}$F 1 hr after peak
temperature).}
\begin{tabular}{ccccccc}
Nova type &M$_{\rm wd}$ (M$_\odot$)
                               & $^{13}$N             & $^{18}$F
          & $^{7}$Be
          & $^{22}$Na
          & $^{26}$Al\\
\hline
CO        & 1.15               & $2.3 \times 10^{-8}$ & $2.6 \times 10^{-9}$
          & $1.1 \times 10^{-10}$
          & $3.8 \times 10^{-12}$
          & $6.2 \times 10^{-10}$\\
ONe       & 1.15               & $2.9 \times 10^{-8}$ & $5.9 \times 10^{-9}$
          & $1.6 \times 10^{-11}$
          & $7.0 \times 10^{-9}$
          & $2.1 \times 10^{-8}$\\
ONe       & 1.25               & $3.8 \times 10^{-8}$ & $4.5 \times 10^{-9}$
          & $1.2 \times 10^{-11}$
          & $6.3 \times 10^{-9}$
          & $1.2 \times 10^{-8}$\\ 
\end{tabular}
\end{table}

\subsection*{Radioactivity in the Galaxy and $\gamma$-ray emission from novae}
An important property of novae ejecta is the presence of radioactive nuclei
(the role of novae as potential $\gamma$-ray emitters was mentioned long ago
\cite{CH74,Cl81,LC87}).

Besides of the very short-lived isotopes responsible of the explosion itself
(see above), there are other short, medium and long-lived nuclei which
have some relevance for the radioactivity of the Galaxy and for the
$\gamma$-ray emission of individual novae. A list of these nuclei with their
main properties is displayed in table 3, whereas the ejected masses of them
obtained from complete hydrodynamical models are shown in table 4. The
short-lived nuclei $^{13}$N and $^{18}$F are produced in similar quantities
in both nova types, whereas $^{7}$Be is mainly produced in CO novae (see
the discussion on $^{7}$Li synthesis above) and $^{22}$Na and $^{26}$Al are
produced in appreciable amounts only in ONe novae. The reason is that in nova
explosions the temperatures reached (around $2-3 \times 10^8$ K, see table 1)
are not high enough to break the CNO cycle; therefore, only if some seed
nuclei (like $^{20}$Ne, $^{23}$Na, $^{24,25}$Mg) are present in the envelope
material, can the NeNa-MgAl cycles operate and synthesize those radioactive
nuclei (and other intermediate-mass isotopes). As CO white dwarfs are devoid
of these nuclei, it is almost impossible for them to produce large
amounts of radioactive $^{22}$Na and $^{26}$Al. 

It is worth mentioning that uncertainties still affect some nuclear reaction
rates of the NeNa-MgAl cycles; this leads to uncertain theoretical
determinations
of the yields of $^{22}$Na and $^{26}$Al, but the error (from a purely nuclear
point of view, and defined as the ratio between maximum and minimum
productions) amounts to factors between 2 and 10 (see \cite{JCH99}
for details). Also the final amount of $^{18}$F synthesized in both CO
and ONe novae is still not well known, since the nuclear reaction rates affecting
$^{18}$F destruction (via $^{18}$F(p,$\alpha$) and $^{18}$F(p,$\gamma$)) are
still not well determined \cite{He99}.

Radioactive nuclei ejected by novae play a role in the radioactivity of the
Galaxy which depends on their lifetimes. The short-lived nuclei (i.e.,
$^{13}$N and $^{18}$F) produce an intense burst of $\gamma$-ray emission,
with duration of some hours, which is emitted before the nova visual
maximum (see \cite{Go98,He99} for details).
This emission is related to positron annihilation, which consists of a line
at 511 keV and a continuum at energies between 20 and 511 keV, related to the
positronium continuum plus the comptonization of the photons emitted in the
line. In figure 2 we show an example of the spectral evolution of a CO and
an ONe nova, at different epochs after peak temperature.

The emission related to
medium-lived nuclei, $^{7}$Be and $^{22}$Na, appears later and is
different in CO and ONe novae, because of their different nucleosynthesis. CO
novae display a line at 478 keV, related to $^{7}$Be decay, whereas ONe novae
show a line at 1275 keV, related to $^{22}$Na decay.

Finally, the long-lived isotope $^{26}$Al is also produced by novae. The
Galactic $\gamma$-ray emission observed at 1809 keV (Mahoney et al.
\cite{Ma82} with the HEAO3 satellite; Diehl et al. \cite{Di95} with the
CGRO/COMPTEL)
corresponds to the decay of $^{26}$Al. Its distribution seems to correspond
better to that of a young population and the contribution of novae is not the
dominant one (see \cite{PD96,JHC97}).

In summary, classical novae explosions produce $\gamma$-rays, being the signature
of CO and ONe novae different. The detectability
distances for the lines at 478 and 1275 keV with the future instrument
INTEGRAL/SPI will range between 0.5 and 2 kpc. To compute these distances,
the width of the lines is taken into account ($\sim$ 7 keV for the 478 keV
line and $\sim$ 20 keV for the 1275 keV line).
The continuum and the 511 keV line are the most intense emissions, but their
appearence before visual maximum and their very short duration requires
``a posteriori" analyses, with monitor-type instruments, with a
large field-of-view and sensitivity up to some hundred keVs. With future
instruments of these characteristics, novae would be detectable more easily
in $\gamma$-rays than visually, because of the lack of extinction.

Future instrumentation in the $\gamma$ and hard X-ray domain will give crucial
insights on the nova theory allowing for a direct confirmation of
the nucleosynthesis in these explosions, but also providing unique information
about the Galactic distribution of novae and their rates.

\end{document}